\documentclass[prd,preprint,tightenlines,floatfix,showpacs,preprintnumbers,nofootinbib,eqsecnum,superscriptaddress]{revtex4}
\usepackage[cp1250]{inputenc} 
\usepackage[T1]{fontenc} 
\usepackage{amsfonts} 
\usepackage{amsopn}   
\usepackage{amssymb}  
\usepackage{amsthm}   
\usepackage{graphicx}
\usepackage{mathtools}
\usepackage{enumerate}
\newcommand{\br}{\mbox{\boldmath $r$}}

\newcommand{\bb}{\mbox{\boldmath $b$}}

\newcommand{\ket}[1]{| {#1} \rangle}
\newcommand{\bra}[1]{\langle {#1} |}

\newcommand{\half}{{1\over 2}}

\def\lsim{\mathrel{\rlap{\lower4pt\hbox{\hskip1pt$\sim$}}
		\raise1pt\hbox{$<$}}}         
\def\gsim{\mathrel{\rlap{\lower4pt\hbox{\hskip1pt$\sim$}}
		\raise1pt\hbox{$>$}}}         

\begin{document}

\vfill
\title{Photoinduced inclusive cross sections in hadronic collisions at the LHC}

\author{Robert G\c{e}barowski}
\email{robert.gebarowski@pk.edu.pl}
\affiliation{
	T.~Kosciuszko Cracow University of Technology, PL-30-084 
	Cracow, Poland}

\author{Agnieszka {\L}uszczak}
\email{agnieszka.luszczak@pk.edu.pl} 
\affiliation{
	T.~Kosciuszko Cracow University of Technology, PL-30-084 
	Cracow, Poland}

\author{Marta {\L}uszczak}
\email{mluszczak@ur.edu.pl} 
\affiliation{
	University of Rzesz\'ow,  
	Rzesz\'ow, Poland}

\author{Wolfgang Sch{\"a}fer}
\email{wolfgang.schafer@ifj.edu.pl} \affiliation{Institute of Nuclear Physics Polish Academy of Sciences, 
	ul. Radzikowskiego 152, PL-31-342 Cracow, Poland}

\date{\today}

\begin{abstract}
We discuss inclusive electromagnetic dissociation processes 
in proton--proton ($pp$) and proton--nucleus ($pA$) scattering at the LHC where one or both of the protons dissociate. These processes which involve the exchange of a virtual photon in the $t$--channel are calculable in terms of deep inelastic structure functions (virtual photoabsorption cross sections). For the $pA \to XA$ reaction there emerges the possibility of measuring the total photoabsorption cross section on the proton at very high energies.
\end{abstract}



\maketitle

\section{Introduction}

Recently there has been much interest in the role of photons in a variety of processes studied at the LHC. Firstly, there is the very active field of ultraperipheral collisions (UPC) of heavy nuclei \cite{Baur:2001jj,Bertulani:2005ru,Klein:2020fmr,Schafer:2020bnm,Bertulani:2023tgh,Grund:2024dhh,Maj:2024dhh} which exploits the large fluxes of coherent Weizs\"acker--Williams photons for highly charged ions.  Secondly, also for protons the importance of treating photons as partons has been realized \cite{Gluck:2002fi,Luszczak:2015aoa,Manohar:2016nzj,Luszczak:2018ntp}.
In this letter we would like to address a few photoinduced processes that have not attracted much attention in the recent literature. These processes do not include production of particles in the central rapidity region via photon--photon fusion but rather involve the (mutual) excitation of colliding particles via photon exchange.
It has been pointed out early \cite{Low:1972ng,Carimalo:1974mw} that these inelastic processes are calculable, with some reservations on the phase space of inelastically excited systems, from deep inelastic structure functions as measured in electron scattering.   


We are interested in the general process $AB \to XY$ mediated by one-photon exchange as shown in the diagrams in Fig.\ref{fig:Diagrams}.
These processes, where either of the colliding particles $A,B$ can be a proton, or heavy nucleus include either mutual excitation with two excited systems of invariant mass $M_X, M_Y$ or excitation of only one of the incoming particles with the other emerging intact in the final state.

One would immediately see that we are talking about the same final state topology as double and single diffractive dissociation, respectively, with a large rapidity gap between final state systems.

\section{Photon exchange cross section from structure functions}

\begin{figure}[t]
\centering
\includegraphics[width=\linewidth]{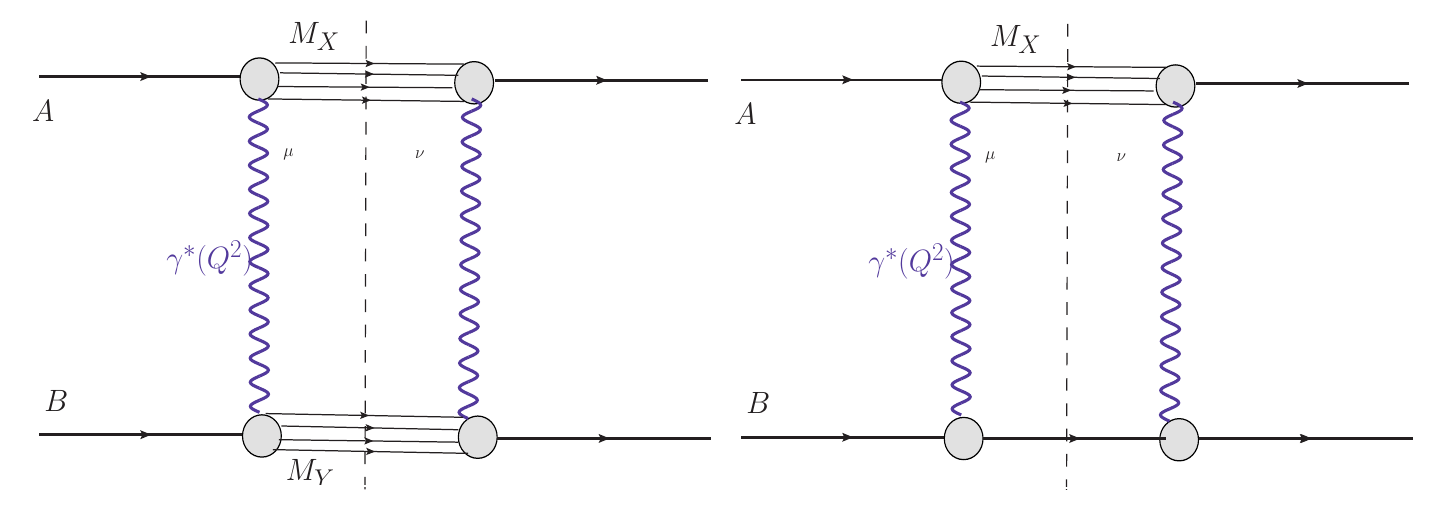}
\caption{Diagrams for the cross sections with dissociation of both hadrons, $AB \to XY$ (left) and one of the hadrons, $AB \to XB$ (right), respectively.}
\label{fig:Diagrams}
\end{figure}


The Feynman diagrams for the cross sections with exchange of a virtual photon  depicted in Fig.\ref{fig:Diagrams}. Here the coupling of photons, say to particle $A$ are given by the so-called hadronic tensor, which is given by the electromagnetic current operator $J_\mu$ integrating over the phase space $d \Phi_X$ of the excited system $X$, which carries four--momentum $p_X$:
\begin{eqnarray}
    W^{(A)}_{\mu \nu}(p_A,q) = \overline\sum (2 \pi)^3 \delta^{(4)}(p_X - p_A -q) 
    \bra{A} J_\mu \ket{X} \bra{X} J^\dagger_\nu \ket{A} \, \, d\Phi_X \, . 
\end{eqnarray}
We decompose the hadronic tensor as
\begin{eqnarray}
W^{(A)}_{\mu \nu}(p_A,q) = - \delta^\perp_{\mu \nu}(p_A,q) \, W_T(M_X^2,Q^2) + e^0_\mu e^0_\nu   \, W_L(M_X^2,Q^2) \, .   
\end{eqnarray}
Here the structure functions $W_{T,L}$ depend on the Lorentz-invariants $M_X^2 = p_X^2$ and $Q^2 = - q^2$.
Labels $T,L$ refer to the transverse and longitudinal  virtual photon polarizations in the frame of the respective hadron, in this case ''$A$''. The polarization vector of the longitudinal photon is 
\begin{eqnarray}
    e_\mu^0 = 2\sqrt{\frac{Q^2}{\lambda(M_X^2,m_A^2,-Q^2)}}\, \Big(p_{A\mu} - \frac{p_A\cdot q}{q^2} q_\mu \Big), \quad e^0 \cdot e^0 = 1.
\end{eqnarray}
Here $\lambda(x,y,z) = x^2 + y^2 + z^2 - 2xy - 2 xz - 2 yz$ is the standard K\"all\'en function, while 
\begin{eqnarray}
\delta_{\mu \nu}^\perp(p_A,q) = g_{\mu \nu} - \frac{q_\mu q_\nu}{q^2} - e^0_\mu e^0_\nu \, . 
\end{eqnarray}
One commonly defines virtual photoabsorption cross sections as
\begin{eqnarray}
    \sigma_T(\gamma^* A; M_X,Q^2) &=& \frac{2 \pi^2 \alpha_{\rm em}}{\phi_A} \frac{-\delta^\perp_{\mu \nu}}{2} \, W^{(A) \mu \nu}(M_X^2,Q^2) \nonumber \\
    \sigma_L(\gamma^*A;M_X,Q^2) &=&  \frac{2 \pi^2 \alpha_{\rm em}}{\phi_A} e_\mu^0 e_\nu^0  W^{(A) \mu \nu}(M_X^2,Q^2) \, . 
\end{eqnarray}
Different conventions are in use for the flux factor $\phi_A$, such as
\begin{eqnarray}
    \phi_A = M_X^2 - m_A^2, \quad {\rm or} \quad  \phi_A = \lambda^{1/2}(M_X^2,m_A^2,-Q^2) \, ,   
\end{eqnarray}
all admissible conventions must reduce to the correct flux factor for real photons ($Q^2 \to 0$), so that the proper real photoabsorption cross section is obtained in the on--shell limit. We will keep $\phi_A,\phi_B$ explicit in our formulas, so that the reader may use the one appropriate to a concrete model/parametrization from the literature.

One often prefers to work with dimensionless structure functions, introduced via
\begin{eqnarray}
\sigma_{T,L}(\gamma^*A;M_X^2,Q^2) = \frac{4 \pi^2 \alpha_{\rm em}}{Q^2} \, \frac{1}{\sqrt{1 + \frac{4 x^2m_A^2}{Q^2}}} \, F_{T,L}(x,Q^2) \, ,     
\end{eqnarray}
with 
\begin{eqnarray}
    x = \frac{Q^2}{M_X^2 + Q^2 - m_p^2} \, .
\end{eqnarray}
The structure function parametrization applies for inelastic processes, $M_X \geq m_p + m_\pi$. 

Now, the cross section for proceses depicted in Fig.\ref{fig:Diagrams}, is obtained from the contraction of two hadron tensors
\begin{eqnarray}
    d\sigma \propto W^{(A)}_{\mu \nu}(p_A,q) \, \frac{1}{Q^4} \, W^{(B)\mu \nu}(p_B,-q) \, .
\end{eqnarray}
When expressed in terms of $\sigma_{T,L}$, the threefold differential cross section can be written in the compact form \cite{Carimalo:1974mw}:
\begin{eqnarray}
 \frac{d \sigma}{d Q^2 dM^2_X dM^2_Y} &=& \frac{1}{16 \pi^3} \,   \frac{\phi_A \phi_B}{Q^4} \Big\{ C_{TT} \, \sigma_T(\gamma^* A; M_X,Q^2) \sigma_T(\gamma^* B; M_Y,Q^2) \, \nonumber \\
 &+& C_{LT} \, \Big( \sigma_T(\gamma^* A; M_X,Q^2) \sigma_L(\gamma^* B; M_Y,Q^2) + \sigma_L(\gamma^* A; M_X,Q^2) \sigma_T(\gamma^* B; M_Y,Q^2) \Big) \nonumber \\
 &+& C_{LL} \, \sigma_L(\gamma^* A; M_X,Q^2) \sigma_L(\gamma^* B; M_Y,Q^2) \Big \}
 \label{eq:master_formula}
\end{eqnarray}
The functions $C_{ij}$ depend on $Q^2, s$ and invariant masses $M_X,M_Y,m_A,m_B$ and are given by
\begin{eqnarray}
    C_{TT} = \frac{1 + \cosh^2 \eta}{\lambda(s,m_A^2,m_B^2)} ,\quad C_{LT} = \frac{\sinh^2 \eta}{\lambda(s,m_A^2,m_B^2)}, \quad C_{LL} = \frac{\cosh^2 \eta}{\lambda(s,m_A^2,m_B^2)} \, , 
\end{eqnarray}
 and
\begin{eqnarray}
   \cosh \eta &=&  \frac{2 Q^2 s}{\lambda^{1/2}(M_X^2,m_A^2,-Q^2)\lambda^{1/2}(M_Y^2,m_B^2,-Q^2)}\, \nonumber \\
   &\times& \Big\{ 1 - \frac{ \Sigma + Q^2}{2s} - \frac{(M_X^2 - m_A^2)(M_Y^2-m_B^2)}{2Q^2s} \Big\} \, .  \nonumber \\
\end{eqnarray}
Notice that the functions $C_{ij}$ become independent of $s$ as $s \to \infty$.
The exact kinematic boundaries for $Q^2$ are
\begin{eqnarray}
    Q^2_{\rm max,  min} = \frac{1}{2} \Big\{ s - \Sigma + \frac{(m_A^2-m_B^2)(M_X^2 - M_Y^2)}{s} \pm \frac{\lambda^{1/2}(s,m_A^2,m_B^2) \lambda^{1/2}(s,M_X^2,M_Y^2)}{s} \Big\} \, ,
\end{eqnarray}
where we have defined
\begin{eqnarray}
    \Sigma = m_A^2 + m_B^2 + M_X^2 + M_Y^2 \, . 
\end{eqnarray}
The inputs to the calculation are the virtual photoabsorption cross section $\sigma_T,\sigma_L$ for incoming hadrons $A$ and $B$. 
Here we note, that the master formula Eq.\ref{eq:master_formula} is fully general in that it also includes the processes with elastic vertices depicted in 
the right panel of Fig.\ref{fig:Diagrams}.
In this case one would obtain
\begin{eqnarray}
   \sigma_T(\gamma^* A; M_X,Q^2) &=& \frac{2 \pi^2 \alpha_{\rm em}}{\phi_A} \, \delta(M_X^2 - m_A^2) \, G_M^2(Q^2)  \nonumber \\
    \sigma_L(\gamma^*A;M_X,Q^2) &=&  \frac{2 \pi^2 \alpha_{\rm em}}{\phi_A} \, \delta(M_X^2 - m_A^2) \, 4 m_A^2 G_E^2(Q^2) \, .
    \label{eq:elastic_cross_sections}
\end{eqnarray}
Note, that the apparent singular (in the limit $M_X^2 \to m_A^2$) factor $1/\phi_A$ cancels in the cross section formula Eq.(\ref{eq:master_formula}). 
Here $G_{M,E}$ are the magnetic and electric Sachs form factors respectively. They reflect the fact, that the longitudinal structure function in the rest frame of the target indeed is related to Coulomb response, whereas the transverse photons induce the magnetic moment interactions. Correspondingly, for the lead nucleus $^{208}Pb$ we keep only the longitudinal part in the elastic piece, while for the proton we include both electric and magnetic formfactors, as one would also do for lighter nuclei as far as the latter are known.

\section{Numerical results}

\subsection{Inelastic--inelastic cross section for $pp$ scattering}

\begin{figure}[t]
\centering
\includegraphics[width=0.49\linewidth]{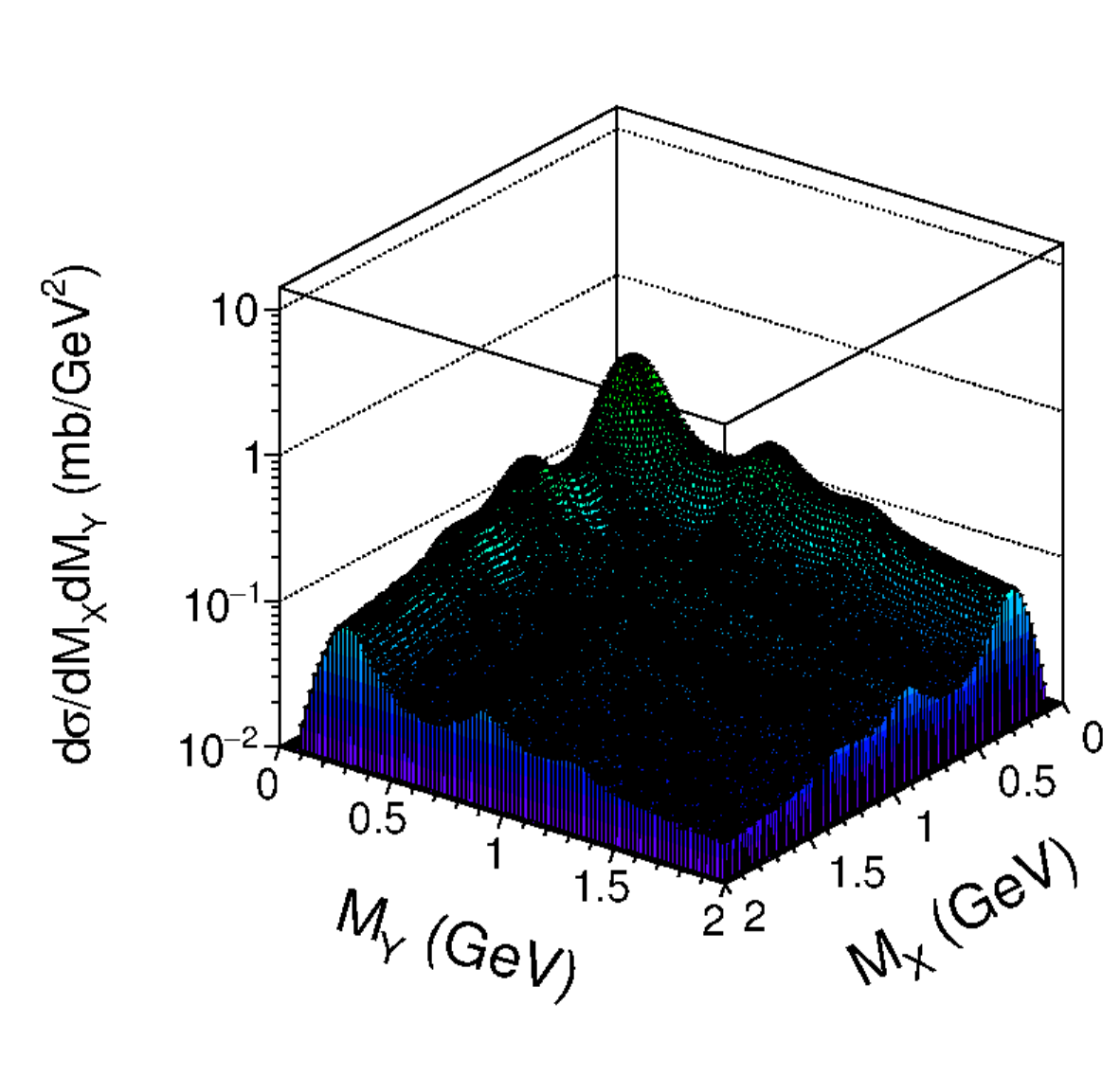}
\includegraphics[width=0.49\linewidth]{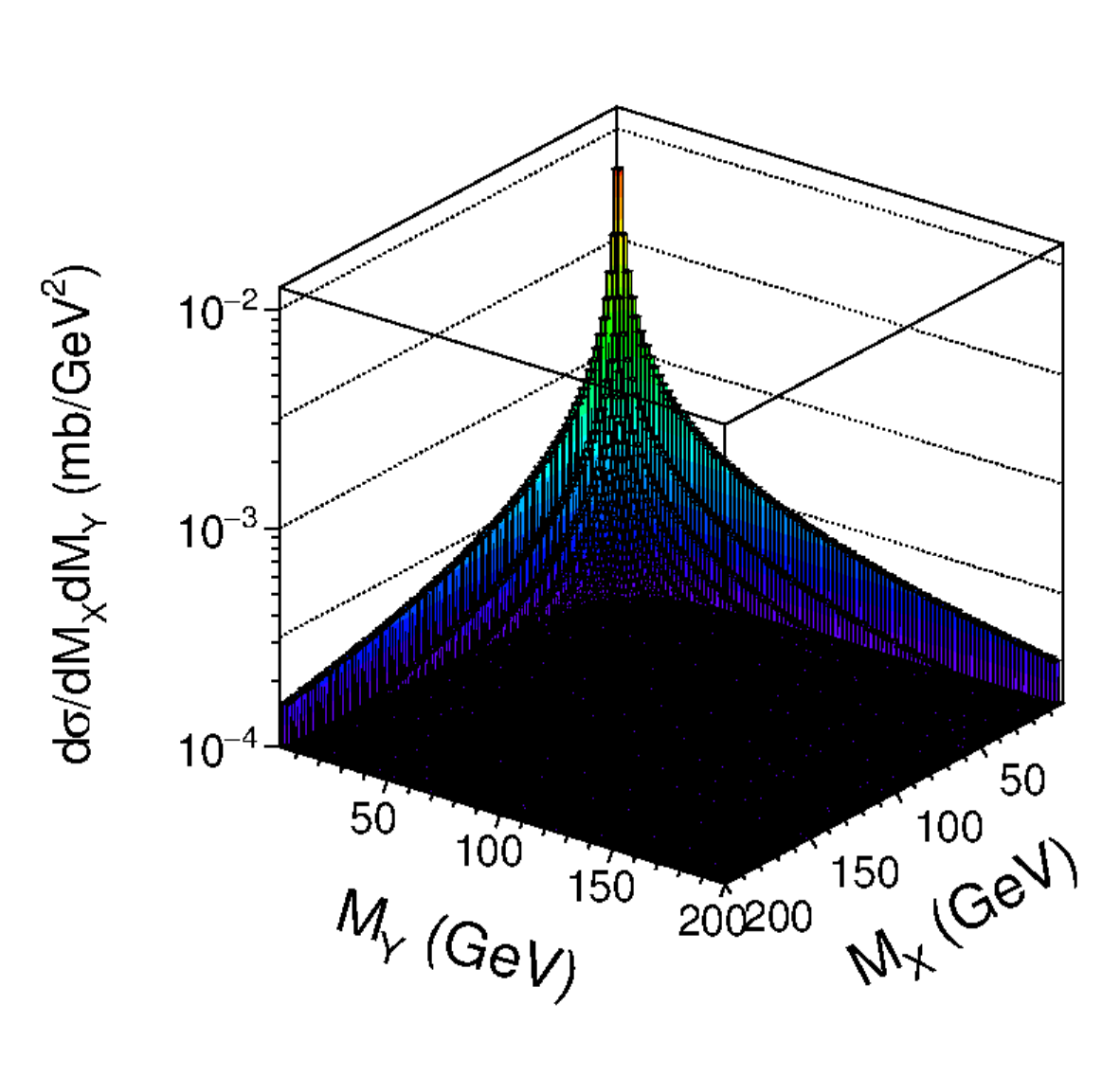}
\caption{Distributions in invariant masses $M_X, M_Y$ for the process $pp \to XY$ at $\sqrt{s} = 13 \, \rm TeV$ via one--photon exchange. The left panel shows the low--mass resonance region of the $\gamma p \to X, \gamma p \to Y$ processes, while the right panel shows large masses, $M_X, M_Y > 2 \, \rm GeV$.}
\label{fig:mapy_mxmy}
\end{figure}
\begin{figure}[t]
\centering
\includegraphics[width=0.49\linewidth]{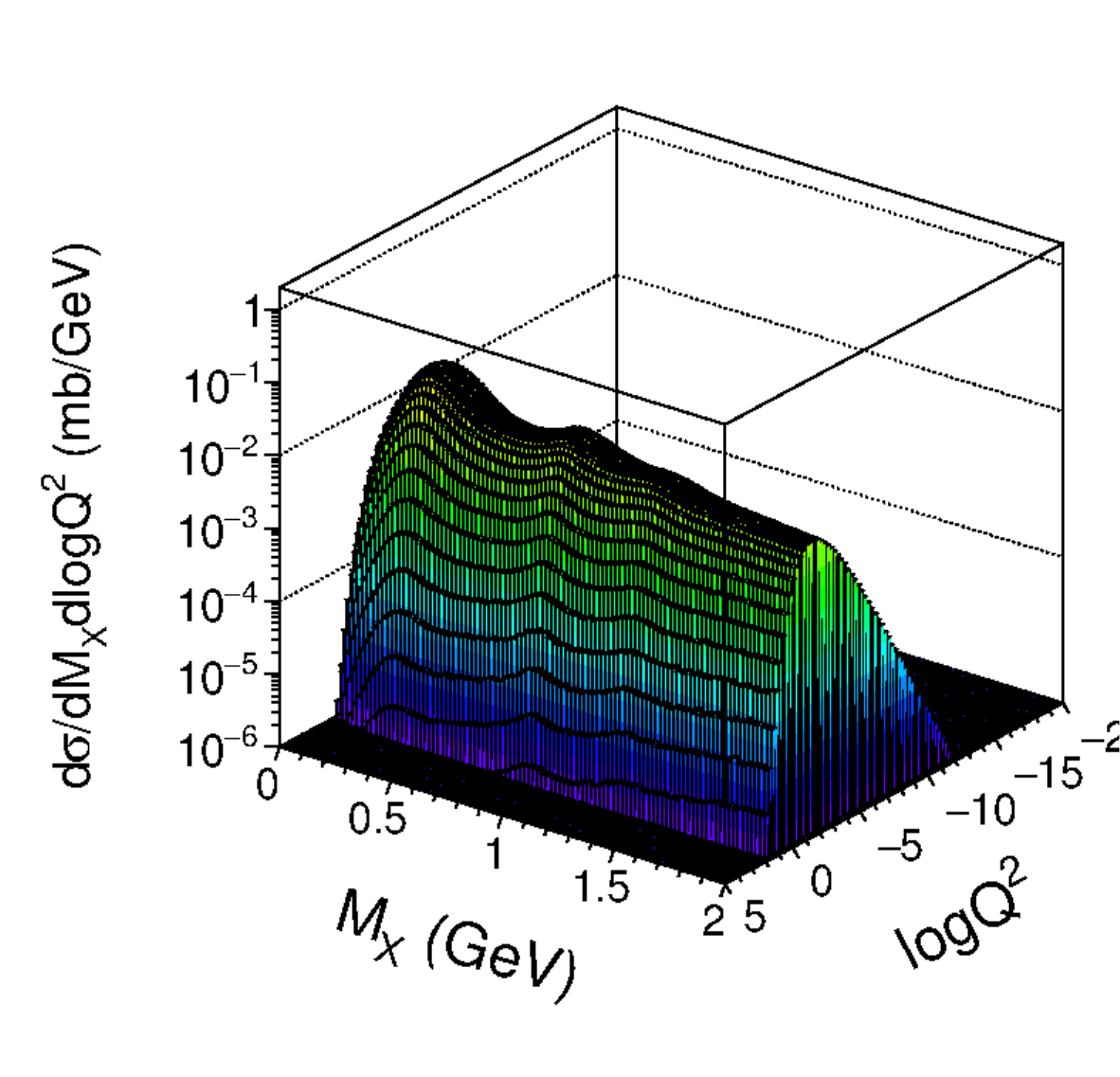}
\includegraphics[width=0.49\linewidth]{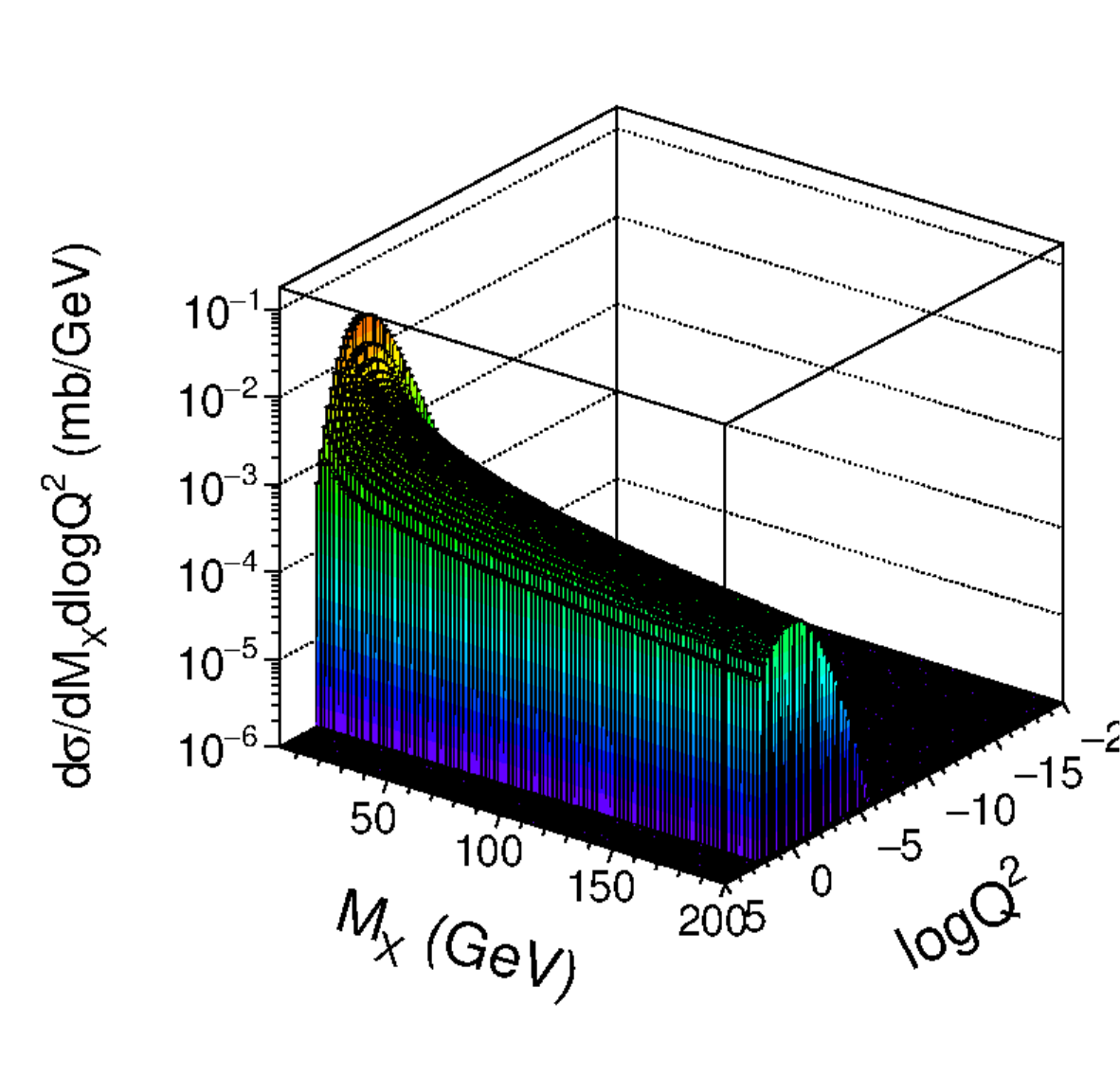}
\caption{Distributions in invariant masses $M_X, M_Y$ for the process $pp \to XY$ at $\sqrt{s} = 13 \, \rm TeV$ via one--photon exchange. The left panel shows the low--mass resonance region of the $\gamma p \to X, \gamma p \to Y$ processes, while the right panel shows large masses, $M_X, M_Y > 2 \, \rm GeV$.}
\label{fig:mapy_mxlogq2}
\end{figure}

We first consider the case where both incoming protons undergo inelastic dissociation through photon exchange, leading to a final state characterized by two hadronic systems $X$ and $Y$ with large rapidity gaps between them. The reaction proceeds via the exchange of a virtual photon, as shown in the left panel of Fig.1, and is computed using the general formalism described in Section II, particularly Eq.(2.10), which expresses the three-fold differential cross section in terms of the virtual photoabsorption cross sections $\sigma_{T,L}(\gamma^*p; M^2, Q^2)$ and kinematic coefficient functions $C_{ij}(s, Q^2, M^2_X, M^2_Y)$.
\begin{figure}[t]
\centering
\includegraphics[width=0.48\linewidth]{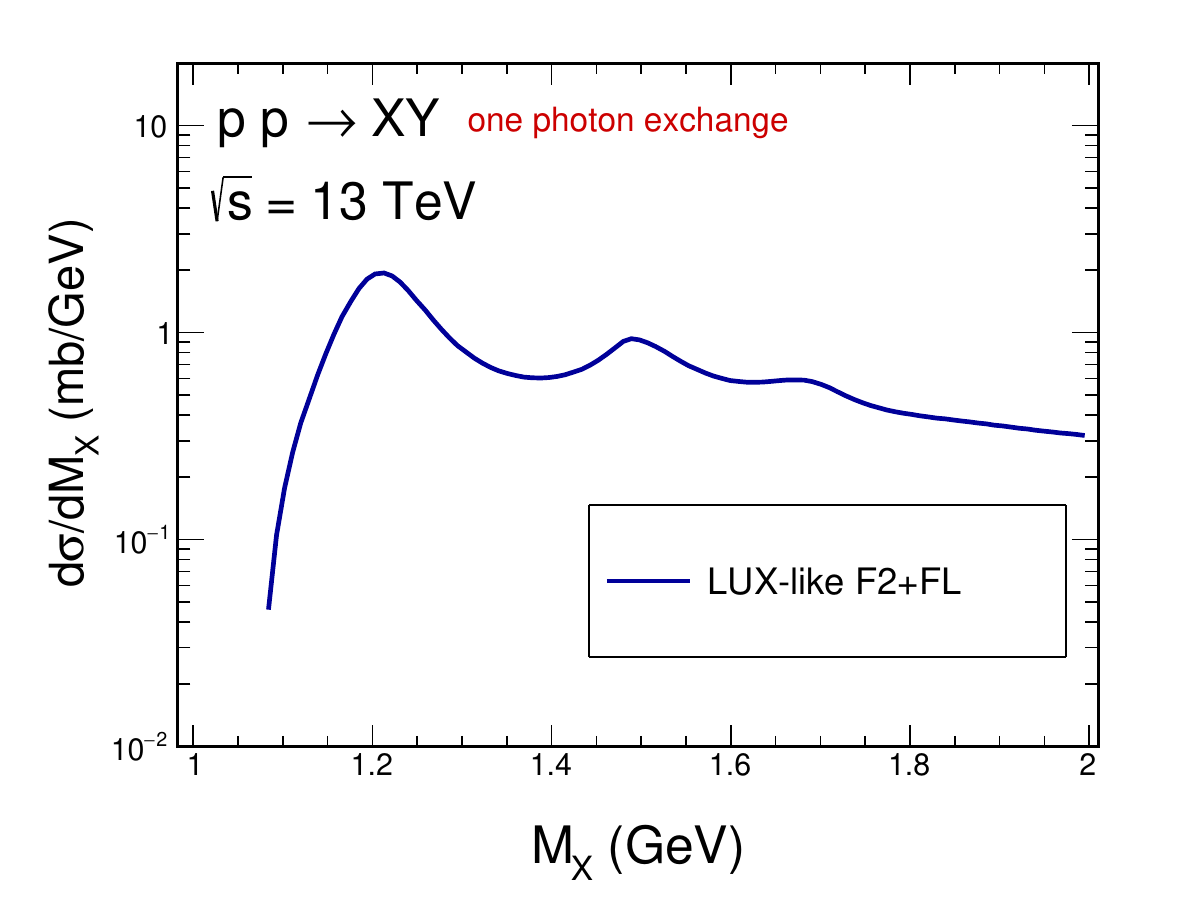}
\includegraphics[width=0.48\linewidth]{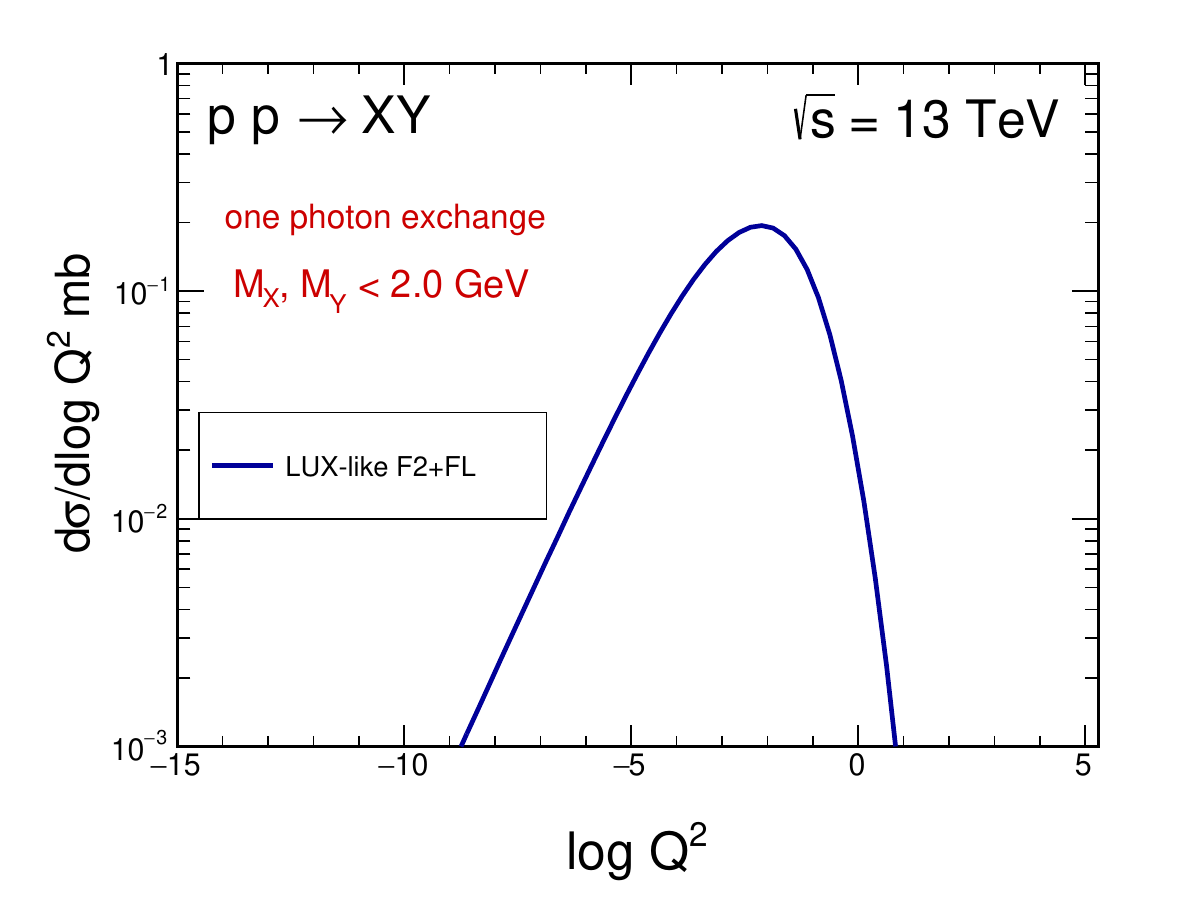}
\includegraphics[width=0.48\linewidth]{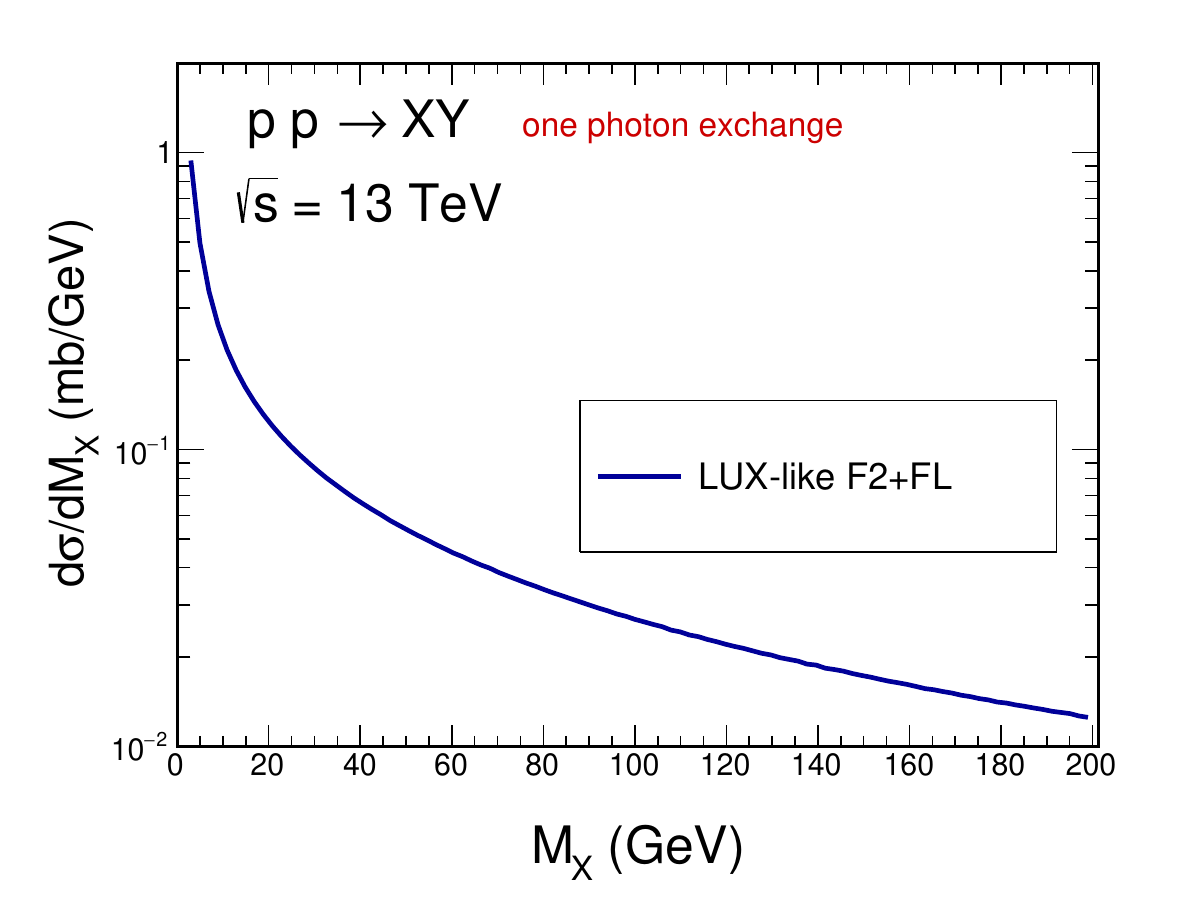}
\includegraphics[width=0.48\linewidth]{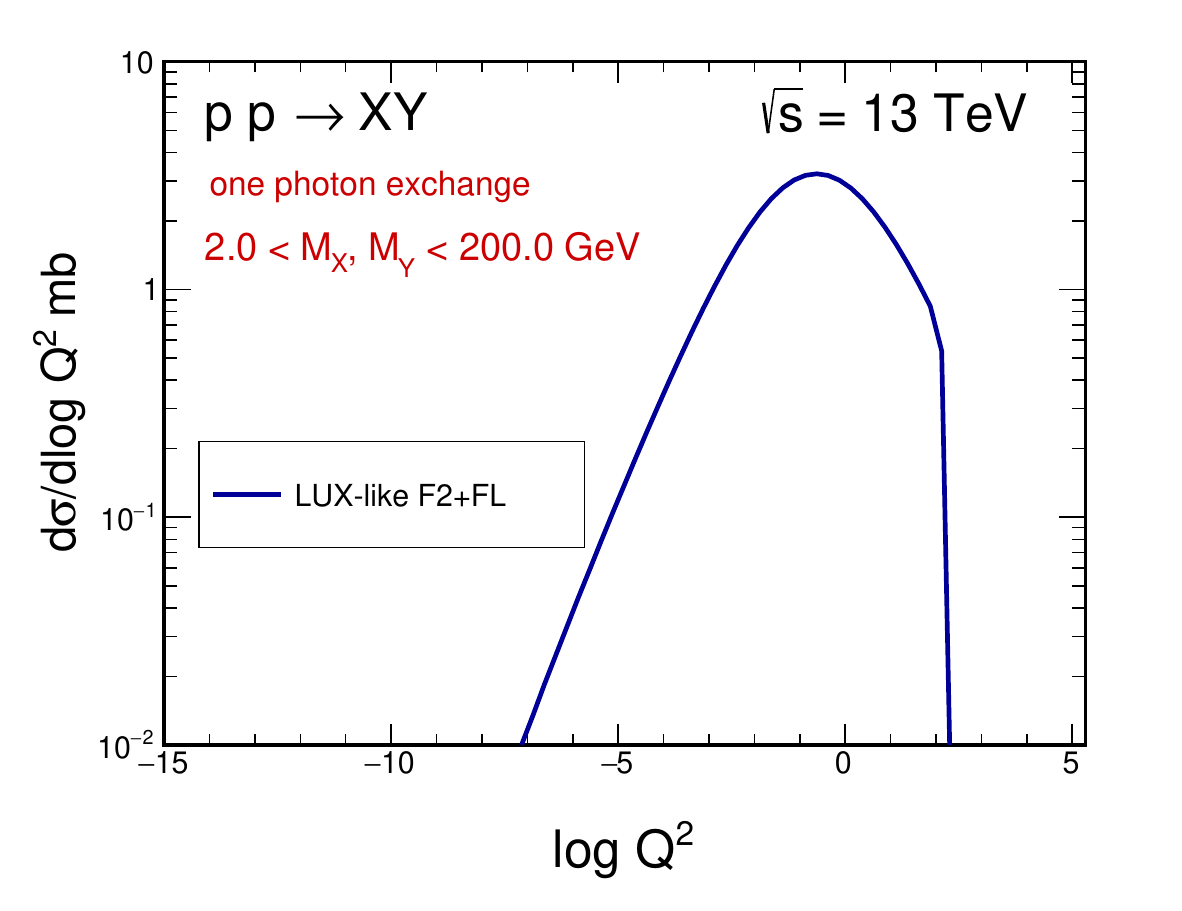}
\caption{Distributions of invariant mass (left panels) and photon virtuality (right panels) in the mutual electromagnetic dissociation of protons, $pp \to XY$. The upper panels are for the low--mass range of $M_X, M_Y<2 \, \rm GeV$, while the right panels show the high--mass contribution, $2 \, {\rm GeV} < M_X, M_Y < 200 \, \rm GeV$.}
\label{fig:proton_diss_inel_inel}
\end{figure}
In our numerical evaluation, we set the proton--proton center-of-mass energy to $\sqrt{s} = 13~\mathrm{TeV}$, relevant for current and future runs of the LHC. The cross sections are computed over a wide range of photon virtualities $Q^2$ and dissociative invariant masses $M_X$ and $M_Y$.
In our calculations we use a parametrization of structure functions that merges a fit of the resonance region \cite{Bosted:2007xd} with the ALLM \cite{Abramowicz:1991xz,Abramowicz:1997ms} parametrization beyond the resonance region. At large photon virtualities also a parametrization in tems of QCD partons is included, corresponding to the ``Lux--like'' fit of \cite{Luszczak:2018ntp} inspired by the procedure proposed in \cite{Manohar:2016nzj}.

In Fig.\ref{fig:mapy_mxmy} we show the two--dimensional distributions in $M_X,M_Y$ both for the low--mass resonance region as well as a larger mass range. We see that the cross section is dominated by the simultaneous excitation of $\Delta^+(1232)$--resonances. The cross section quickly drops with invariant masses of the excited system. We see that large mass on one arm is strongly correlated with small--mass excitation on the other side.
In Fig.\ref{fig:mapy_mxlogq2} we show two dimensional maps in the invariant mass $M_X$ of one of the dissociated systems and (the logarithm of) $Q^2$. Smaller masses are associated with smaller $Q^2$, that is clearly correlated with the dependence of $Q^2_{\rm min}$
with $M_X,M_Y$. This is also illustrated by Fig.\ref{fig:proton_diss_inel_inel}, where we show one--dimensional distributions in one invariant mass as well as $Q^2$, again for low--mass regions and high--mass regions separately.
\subsection{Elastic--inelastic processes}
\begin{figure}[t]
\centering
\includegraphics[width=0.7\linewidth]{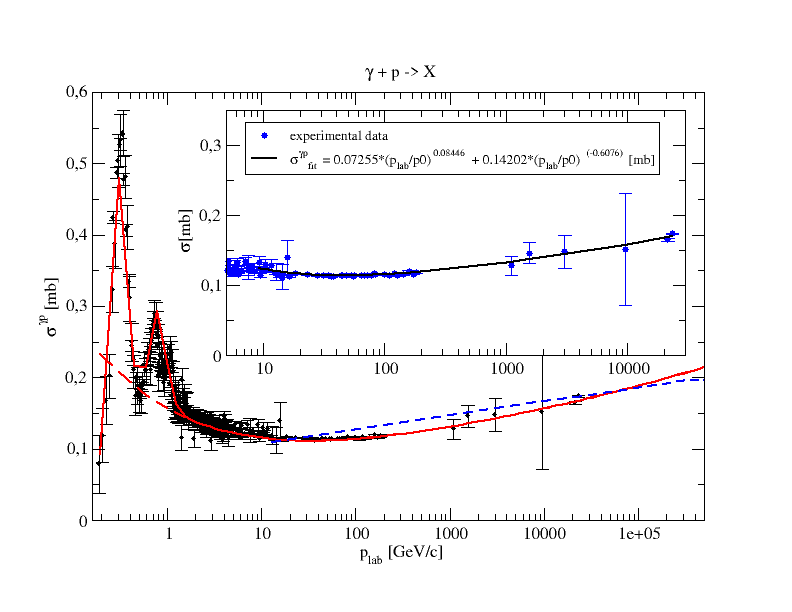}
\caption{Total $\gamma p$ photoabsorption cross section. Data are from \cite{Armstrong:1971ns,Bloom:1969pn,Michalowski:1977eg,Meyer:1970fya,Hilpert:1968uzd,Perl:1969kf,Ballam:1971yd,Caldwell:1978yb,Vereshkov:2003cp,H1:1995hmw,ZEUS:2001wan}. 
The red curve merges a fit from \cite{Bosted:2007xd} with the ALLM parametrization \cite{Abramowicz:1991xz,Abramowicz:1997ms} being the low--$Q^2$ limit of the Lux-like fit in Ref.\cite{Luszczak:2018ntp}. The red dashed line is the ALLM curve alone. The blue dashed line is a color--dipole model calculation (see text).
The inlay shows a power--law fit for the high--energy domain.}
\label{fig:sigma_gamma_p_data}
\end{figure}
The elastic--inelastic processes are of great interest as they could be utilized to measure the total photoabsorption cross section of the dissociating target. As the photon flux emitted by the ``intact'' hadron shows a broad--band energy spectrum, such a determination of a cross section at fixed $\gamma$--hadron center--of--mass energy would require a measurement of the full dissociative hadronic final state and calorimetric determination of its invariant mass.
As for heavy nuclei the longitudinal response proportional to the square of the charge form factor dominates, we can anticipate the dominance of quasi--real photons.
We show experimental data \cite{Armstrong:1971ns,Bloom:1969pn,Michalowski:1977eg,Meyer:1970fya,Hilpert:1968uzd,Perl:1969kf,Ballam:1971yd,Caldwell:1978yb,Vereshkov:2003cp,H1:1995hmw,ZEUS:2001wan} on the total photon--proton cross section in Fig.\ref{fig:sigma_gamma_p_data}. Here the low--energy region studied since the late 1960s  mainly is of interest for the investigation of nucleon resonances, see for example the recent review article \cite{Thiel:2022xtb}. The high--energy region is of interest in its own right and in the context of the QCD--Pomeron exchange.
Here, the transition from deep inelastic structure functions to real photoabsorption allows to investigate the transition from hard (short--distance) to soft (long--distance) hadronic interactions.
A number of theoretical approaches have been suggested, see for example \cite{Nikolaev:1999qh,Watanabe:2012uc,Block:2014kza,Britzger:2019lvc}.
In the high--energy region with exception of the last two data points, data are derived from photoabsorption on nuclei in the interaction of cosmic--ray muons with scintillation detectors \cite{Vereshkov:2003cp}. The highest energy measurements at $W \sim 200 \, \rm GeV$ were performed at the HERA accelerator by the H1 \cite{H1:1995hmw} and ZEUS \cite{ZEUS:2001wan} collaborations, for a review see \cite{Chwastowski:2003aw}. 
For illustration, we show in Fig.\ref{fig:sigma_gamma_p_data} three model curves for the total photoabsorption cross section. First, the solid red curve shows the ``Lux--like" fit of Ref.\cite{Luszczak:2018ntp}, while the red dashed curve is the ALLM fit \cite{Abramowicz:1991xz,Abramowicz:1997ms}. We also show a calculation based on the color dipole approach
\cite{Nikolaev:1990ja}, where the total absorption cross section can be written as
\begin{eqnarray}
    \sigma_T(\gamma^* p; W^2) = \sum_{f =u,d,s,c} \int \frac{dr}{r} \, W_{f \bar f} (r) \, \sigma(x_{f \bar f},r) \, ,
\end{eqnarray}
with 
\begin{eqnarray}
 W_{f \bar f} (r) = \frac{2 \alpha_{\rm em}}{ \pi} e_f^2  \, (m_f r)^2 K_1^2(m_f r) 
\end{eqnarray}
Here for the quark masses we take \cite{Luszczak:2016bxd}
\begin{eqnarray}
    m_u = m_d = m_s = 140 \, {\rm MeV}, m_c = 1.5 \, {\rm GeV} .
\end{eqnarray}
The dipole cross section becomes dependent on the $\gamma p$ energy via
\begin{eqnarray}
    x_{u \bar u} = x_{d \bar d}= \frac{m_\rho^2}{W^2 }, \,  x_{s \bar s} = \frac{m_\phi^2}{W^2}, \, x_{c \bar c} = \frac{4 m_c^2}{W^2} .
\end{eqnarray}
We use a fit to the dipole cross section of Ref.\cite{Luszczak:2016bxd} based on DIS data for $Q^2 > 3.5 \, {\rm{GeV}}^2$. The fact that it decribes the HERA data by magnitude may be somewhat fortuituous. We illustrate with it the energy dependence which comes from the admixture of small dipoles probing the gluon distribution of the proton which drives the energy dependence of the cross section. We see that this dynamically generated energy dependence appears to be somewhat slower than the one in the Regge model fit.
\begin{figure}[t]
\centering
\includegraphics[width=0.48\linewidth]{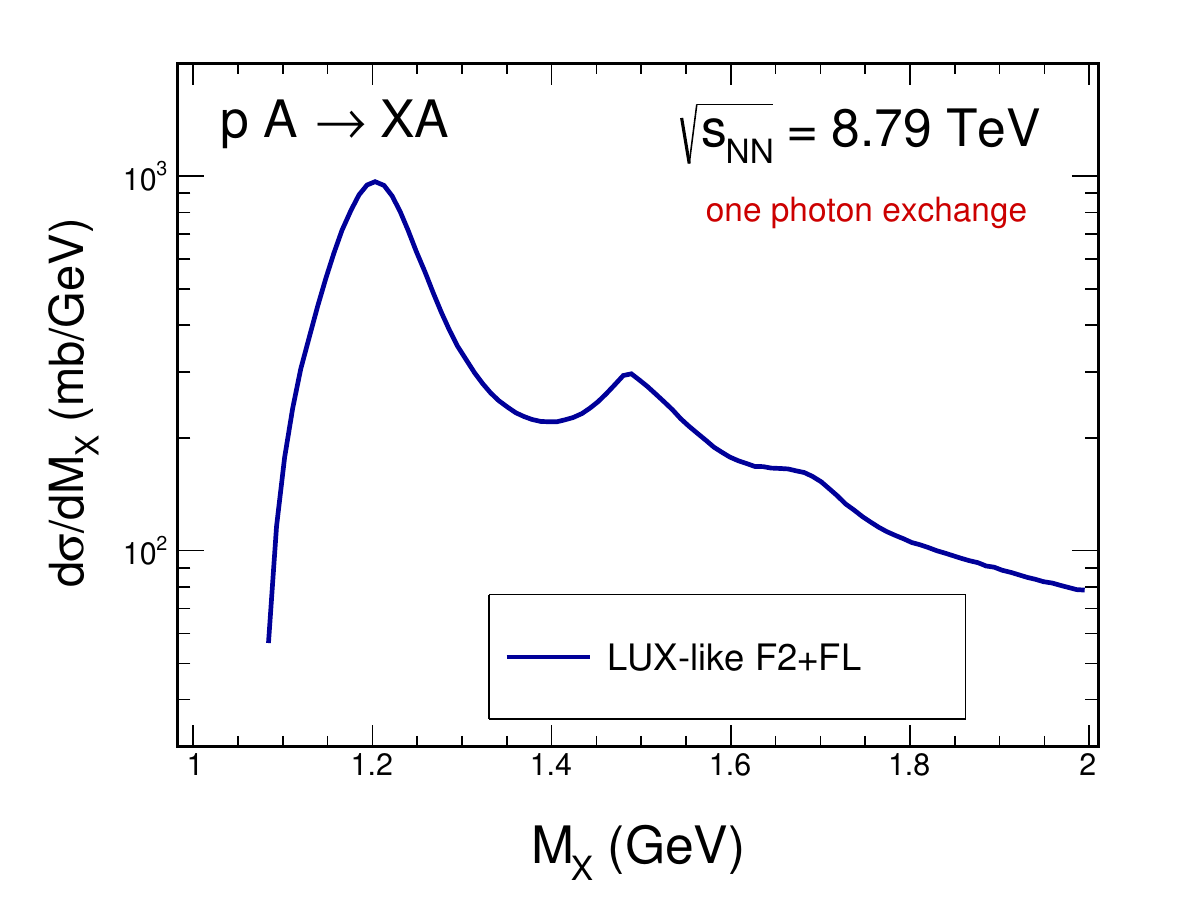}
\includegraphics[width=0.48\linewidth]{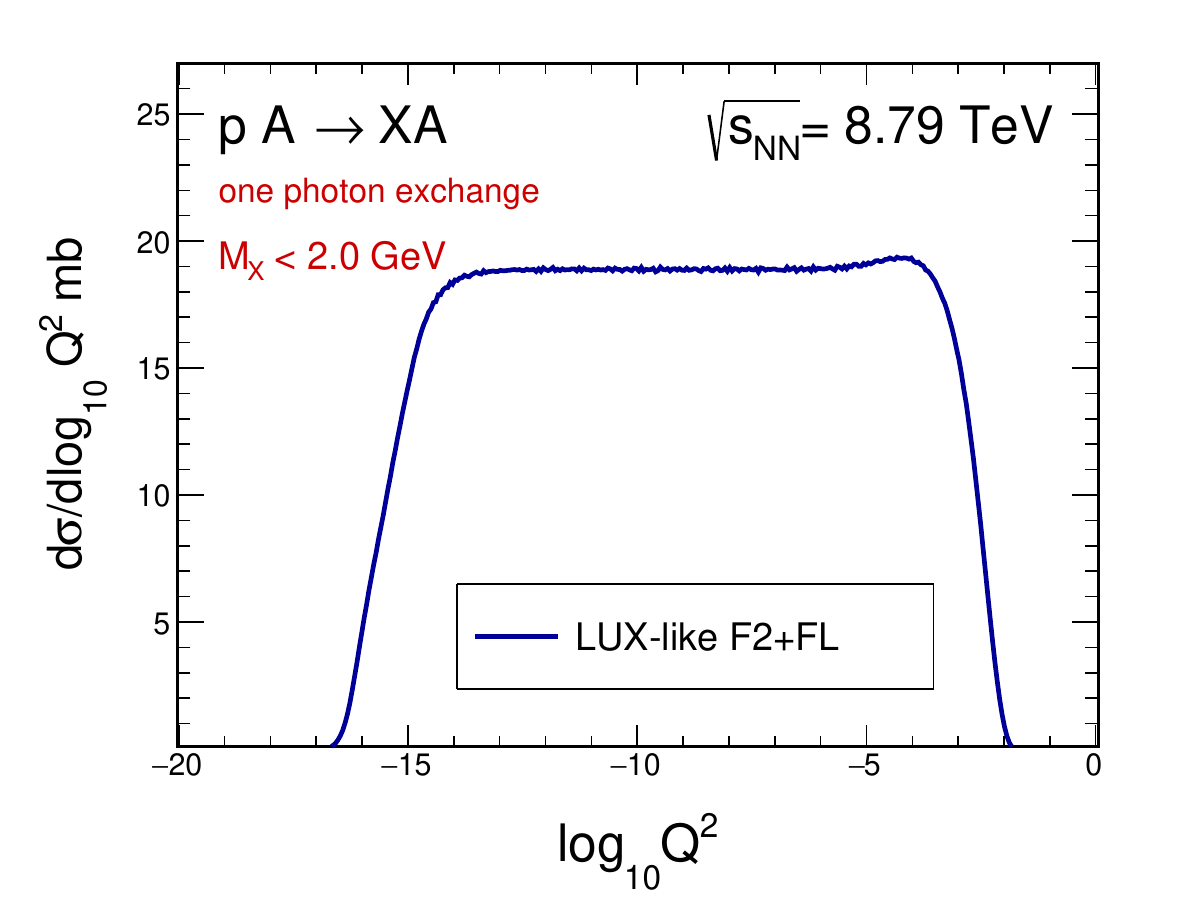}
\includegraphics[width=0.48\linewidth]{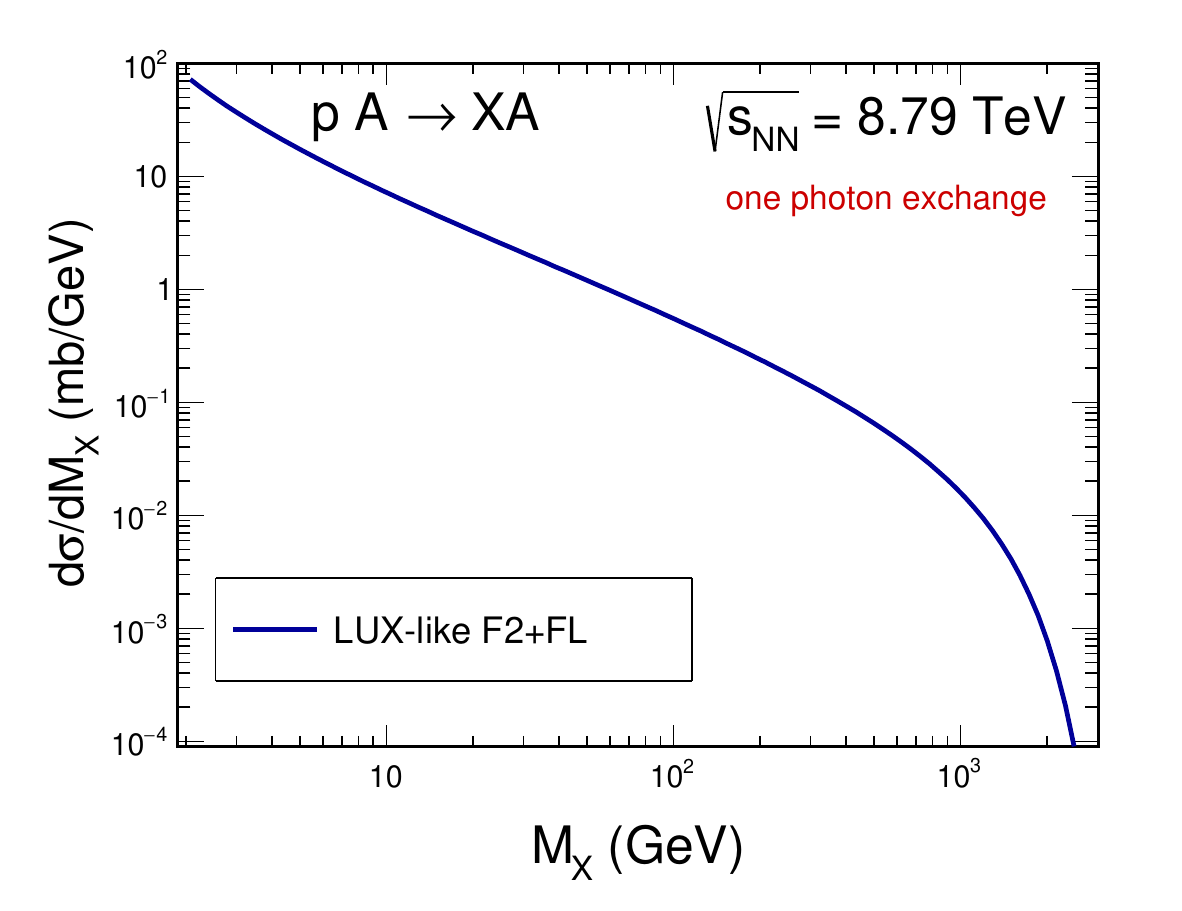}
\includegraphics[width=0.48\linewidth]{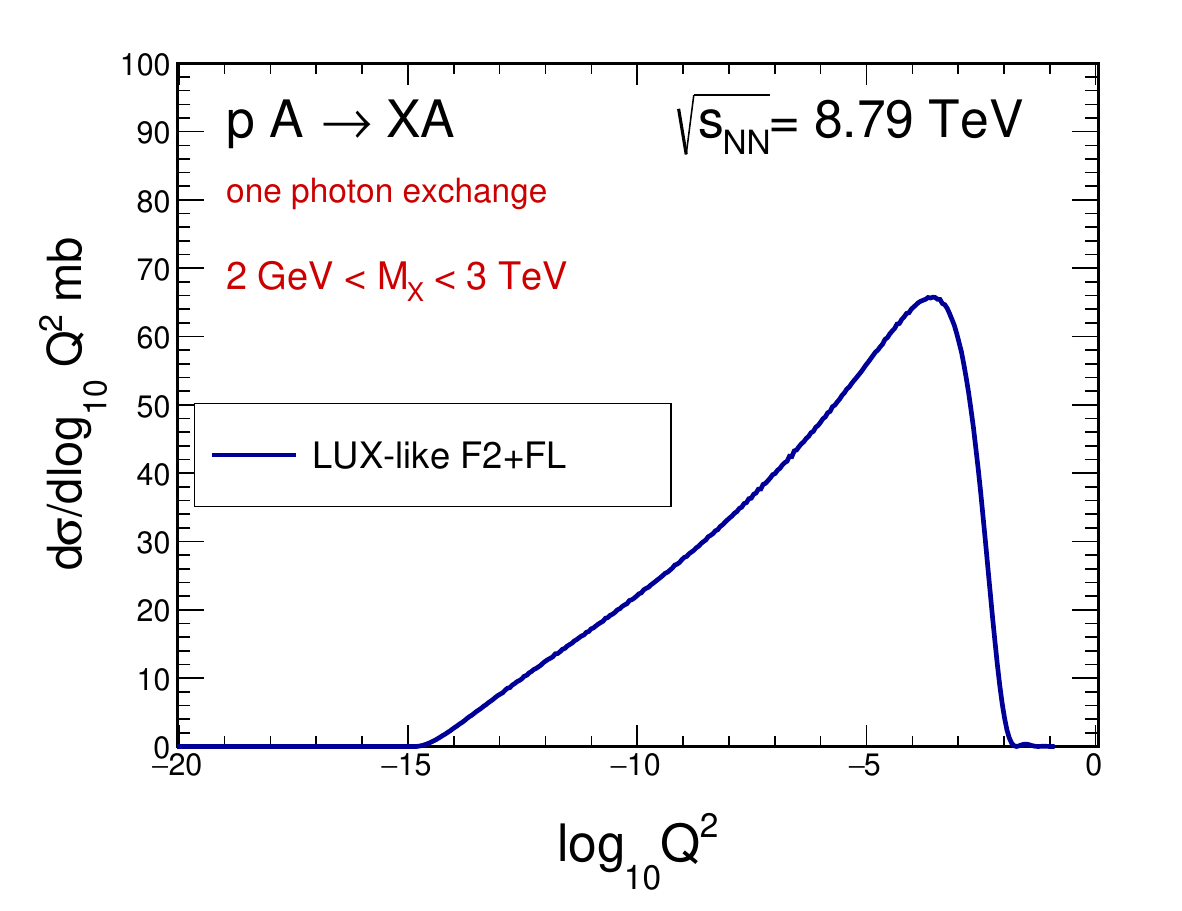}
\caption{Distributions of invariant mass (left panels) and photon virtuality (right panels) in the electromagnetic dissociation of protons on nuclei. The upper panels are for the low--mass range of $M_X<2 \, \rm GeV$, while the right panels show the high--mass contribution.}
\label{fig:proton_diss_on_nuclei}
\end{figure}
We now turn to the process to which one may also refer to as a Coulomb dissociation of protons on the nuclear target.  In Fig.\ref{fig:proton_diss_on_nuclei} we show the differential cross sections for protons colliding withe $^{208}Pb$ ions. In the low--mass region we see the texbook-like $dQ^2/Q^2$ plateau indicating the Weizs\"acker--Williams flux of quasireal photons. 
In the high--mass region the $Q^2$--dependence reflects the rise of $Q_{\rm min}^2$ with $M_X$.
While our calculation includes also $F_L$, its contribution is generally negligible. We see that sizeable cross sections can be obtained out to the TeV-region of $\gamma p$ cm--energy.




\section{Conclusions}
We have presented the calculation of inclusive inelastic cross sections in hadron--hadron collisions induced by $t$--channel photon exchange. These processes have the same event
topology as (double--)diffractive dissociation and can be referred to as a (mutual) Coulomb dissociation. Their distinctive feature is a dominance of very small four--momentum transfers $-t = Q^2$, close to the kinematical limit of the respective process.
Further studies are necessary to assess the potential of measuring the total photoabsorption cross section in proton--nucleus collisions.

\acknowledgements
W.S. would like to thank  Rainer Schicker for an invitation to the EMMI Rapid Reaction Taskforce meeting on ``Next Generation Facility for Forward Physics" in Heidelberg, February 2025, and its participants for discussions, which inspired the present note.
This work was supported by 
the Polish National Science Center Grant No. UMO-2023/49/B/ST2/03665.

\bibliography{gammainduced}

\end{document}